\documentclass[10pt,
    journal,
    compsoc,
    showpacs,
    reprint,
    twocolumn,
    superscriptaddress,
    floatfix,
    amsmath
]{IEEEtran}

\ifCLASSOPTIONcompsoc
  \usepackage[nocompress]{cite}
\else
  \usepackage{cite}
\fi

\usepackage[font=small,labelfont=bf]{caption}
\usepackage{braket}

\usepackage{graphics}
\usepackage{multirow}
\usepackage{dcolumn}
\newcolumntype{.}{D{.}{.}{-1}}

\usepackage{epsfig}
\usepackage{graphicx}
\usepackage{epstopdf}
\usepackage{hyperref}
\usepackage{verbatim}
\usepackage[english]{babel}

\usepackage{color}
\usepackage{listings}
\definecolor{codegreen}{rgb}{0,0.6,0}
\definecolor{codegray}{rgb}{0.5,0.5,0.5}
\definecolor{codepurple}{rgb}{0.58,0,0.82}
\definecolor{backcolour}{rgb}{0.98,0.98,0.98}
 
\lstdefinestyle{mystyle}{
    backgroundcolor=\color{backcolour},   
    commentstyle=\color{codegreen},
    keywordstyle=\color{magenta},
    numberstyle=\tiny\color{codegray},
    stringstyle=\color{codepurple},
    basicstyle=\ttfamily\scriptsize,
    breakatwhitespace=false,         
    breaklines=true,                 
    captionpos=b,                    
    keepspaces=true,                 
    numbersep=5pt,                  
    showspaces=false,                
    showstringspaces=false,
    showtabs=false,                  
    tabsize=2
}
 
\begin{document}

\title{Comparative benchmarking of cloud computing vendors with High Performance Linpack}

\author{
    \IEEEauthorblockN{
        Mohammad Mohammadi, Timur Bazhirov}\\
        \IEEEauthorblockA{
            Exabyte Inc., San Francisco, California 94103, USA\\
        }
}

\IEEEtitleabstractindextext{%
\begin{abstract}

    We present a comparative analysis of the maximum performance achieved by the Linpack benchmark\cite{2003-dongarra-hpl-explained} on compute intensive hardware publicly available from multiple cloud providers. We study both performance within a single compute node, and speedup for distributed memory calculations with up to 32 nodes or at least 512 computing cores. We distinguish between hyper-threaded and non-hyper-threaded scenarios and estimate the performance per single computing core. We also compare results with a traditional supercomputing system for reference. Our findings provide a way to rank the cloud providers and demonstrate the viability of the cloud for high performance computing applications.

\end{abstract}

}

\maketitle

\section{Introduction}
\label{sec:introduction}

    During the last decade cloud computing established itself as a viable alternative to on-premises hardware for mission-critical applications in multiple areas \cite{2009-cloud-computing-definition-nist-mell, 2010-cloud-computing-overview-zaharia, 2016-cloud-computing-overview-rittinghouse}. For high performance computing (HPC)
    workloads that traditionally required large and cost-intensive hardware procurement, however, the feasibility and advantages of cloud computing are still debated. In particular, it is often questioned whether software applications that require distributed memory can be efficiently run on "commodity" compute infrastructure publicly available from cloud computing vendors \cite{2010-jackson-cholia-lbl-cloud-con}.
    
    Several studies reported on the poor applicability of cloud-based environments for scientific computing. Multiple research groups ran both standard benchmark suites such as Linpack and NAS \cite{r-masud, 2008-ostermann, 2008-walker}, and network performance tests \cite{2010-wang}. The cost of solving a system of linear equations was found to increase exponentially with the problem size, illustrating that cloud was not mature enough for such workloads in \cite{2009-napper}. A study of the impact of virtualization on network performance reported significant throughput instability and abnormal delay variations \cite{2010-wang}. An empirical performance evaluation attempted in \cite{2011-iosup} found that while cloud computing services are insufficient for scientific applications at large, they may still be a good solution for the scientists who need resources instantly and temporarily. The performance of a set of typical scientific supercomputing workloads on Amazon EC2 was also found to be lower than for traditional HPC systems in \cite{2010-jackson-cholia-lbl-cloud-con, 2011-cloud-computing-magellan-report}.
    
    Some prior studies had a positive view on the use of cloud computing. The performance of selected bio-informatics and astronomy software was examined, and cloud was found to provide a feasible, cost-effective model in \cite{2008-hazelhurst, 2008-deelman}. In \cite{2008-evangelinos} it was found that cloud is capable of supporting responsive on-demand, small sized HPC applications. The evaluation of micro-benchmarks, kernels, and e-Science workloads in \cite{2008-ostermann} found low performance and reliability, however reported on the potential applicability for scientists that need resources immediately and temporarily. The costs and challenges associated with running a diverse set of science applications on the cloud were studied and found to hold promise in \cite{2007-keahey, 2008-keahey, 2009-keahey, 2010-li, 2010-ramakrishnan}. In \cite{2010-rehr} it was shown that Amazon Elastic Compute Cloud is a feasible platform for applications that do not require advanced network performance. A general review of the field of HPC applications and their state in cloud computing was also conducted in \cite{2012-sanjay}.

    Recent advancements have made it possible to access large-scale computational resources completely on-demand in a rapid and efficient manner. When combined with high fidelity simulations, they can serve as an alternative pathway to enable computational discovery and design of new materials through high-throughput screening. At Exabyte Inc. we have previously demonstrated this with a case study involving high-throughput screening of structural alloys using modeling tools rooted in first-principles quantum mechanical techniques \cite{2016-exabyte-aps-abstract}. During an example run we were able to scale to 10,656 computing cores within 7 minutes from the start. This motivated the need for further benchmarking. In order to address the concerns about the specificity of the materials simulation techniques employed during the aforementioned case study we decided to use a more general tool for the purpose of our current analysis.
    
    In this work we benchmark the performance of the publicly available cloud computing hardware with High Performance Linpack \cite{hpl-top500, 2003-dongarra-hpl-explained}, the benchmark that during the last two decades was employed to rank the top supercomputing systems \cite{hpl-top500} on the global scale. We compare 4 cloud computing vendors, and include results for a traditional supercomputer (number 60 on the top500.org list at the moment of this writing \cite{top500-edison-supercomputer}). Our findings demonstrate that the best-in-class cloud computing options can already deliver similar scaling patterns and match, if not exceed, the performance per core of the more traditional high performance computing systems.

\section{Methodology}
\label{sec:methodology}

    Benchmarking presented in this article is done through High Performance Linpack (HPL). The program solves a random system of linear equations, represented by a dense matrix, in double precision (64 bits) arithmetic on distributed-memory computers. It does so through a two-dimensional block-cyclic data distribution, and right-looking variant of the LU factorization with row partial pivoting. It is a portable and freely available software package. HPL provides testing and timing means to quantify the accuracy of the obtained solution as well as the time-to-completion. The best performance achievable depends on a variety of factors, and the algorithm is scalable such that its parallel efficiency is kept constant with respect to per processor memory usage. Readers may consult the following references for more information: \cite{2003-dongarra-hpl-explained, hpl-netlib, hpl-top500, linpack-userguide}.
    
    Below we present the content of an example input file for the HPL benchmark suite. Ns is matrix size for the underlying system of linear equations, Ps and Qs are the process grid dimensions. These parameters are changed for each reported case based on the number of cores and memory used. In order to achieve the optimal performance, the largest problem size that fits in the memory should be selected. The amount of memory used by HPL is dependent on the size of the coefficient matrix. The logic behind choosing the exact input parameters could be demonstrated by the following line of thought. In case of 4 nodes with 256 Mb of memory each, there is a total of 1 Gb, or 125 million double precision (8 bytes) elements. The square root of this number is 11585. As one has to leave memory for the operating system as well as for other system processes, so a problem size of 10000 would be a good fit. Ps and Qs depend on the physical interconnection network. As a rule of thumb P and Q are taken to be approximately equal, with Q slightly larger than P.


\begin{lstlisting}
    HPL.out      output file name
    6            device out
    1            # of problems
    456768       Ns
    1            # of NBs
    192          NBs
    1            PMAP process mapping
    1            # of process grids (P x Q)
    32           Ps
    36           Qs
    16.0         threshold
    1            # of panel fact
    1            PFACTs
    1            # of recursive stopping criterion
    4            NBMINs
    1            # of panels in recursion
    2            NDIVs
    1            # of recursive panel facts
    1            RFACTs
    1            # of broadcast
    6            BCASTs
    1            # of lookahead depths
    0            DEPTHs
    0            SWAP
    1            swapping threshold
    1            L1
    1            U
    0            Equilibration
\end{lstlisting}

\section{Results}
\label{sec:results}

     We present the cloud server instance types and hardware specification for all studied cases inside Table \ref{table:Hardware}. We choose the highest performing servers available in an on-demand fashion. Most of the compute servers have 16 physical cores and all have at least 2GB per or random access memory per core. The network options differ quite a bit, from 54 to 1 gigabit per second in bandwidth. We also provide metrics for the traditional supercomputing system used as reference \cite{top500-edison-supercomputer}.

\begin{table}
    \caption{
    Hardware specification for the compute nodes used during benchmarking. Core count for physical computing cores and processor frequency, in GHz, are given together with Memory (RAM) size, in gigabytes, and network bandwidth in gigabit-per-second \cite{aws-instance-types, azure-instance-types, rackspace-instance-types, softlayer-instance-types}.
    }
    \label{table:Hardware}
    \begin{tabular}{cccccc}
        Provider & Nodes & Cores & Freq. & RAM & Net\\
        \hline
        AWS-*       & c4.8xlarage       & 18 & 2.9 & 60  & 10 \\
        Azure-AZ    & Standard\_F16s    & 16 & 2.4 & 32  & 10 \\
        Azure-IB-A  & A9                & 16 & 2.6 & 112 & 32 \\
        Azure-IB-H  & H16               & 16 & 3.2 & 112 & 54 \\
        SoftLayer   & Virtual           & 16 & 2.0 & 64  & 1 \\
        Rackspace   & Compute1-60       & 16 & 2.8 & 60  & 5 \\
        NERSC       & Edison            & 24 & 2.4 & 64  & 64
    \end{tabular}
\end{table}

\subsection{Amazon Web Services}
\label{subsec:AWS}

    For Amazon Web Services (AWS) we study 3 different scenarios: the default hyper-threaded, non-hyper-threaded and non-hyper-threaded mode with placement group option enabled. The c4.8xlarge instance types are used.

\subsubsection{Hyper-threaded regime}

    Table \ref{table:AWS} shows the results for AWS instances with hyper-threading enabled (default regime). It can be seen that the ratio of absolute speedup to the number of nodes rapidly decreases as the node count increases.

\begin{table}
    \caption{
    [AWS] Results for Amazon Web Services c4.8xlarge instances with hyperthreading enabled (default scenario). Core count is given for virtual (hyper-threaded) computing cores. Numbers of computing nodes (Nodes) and total computing cores (Cores) are given together with the maximum achieved (Rmax) and peak (Rpeak) performance indicators, and the absolute achieved speedup (Speedup). It can be seen that the ratio of absolute speedup to the number of nodes falls rapidly as the number of nodes is increased.
    }
    \label{table:AWS}
    \begin{tabular}{ccccc}
        Nodes & Cores & Rmax (TFLOPS) & Rpeak (TFLOPS) & Speedup \\
        \hline
        1 & 36 & 0.53 & 1.63 & 1.00 \\
        2 & 72 & 0.98 & 3.26 & 1.85 \\
        4 & 144 & 1.51 & 6.53 & 2.87 \\
        8 & 288 & 2.90 & 13.05 & 5.50 \\
        16 & 576 & 5.23 & 26.10 & 9.92 \\
        32 & 1152 & 8.65 & 52.20 & 16.41
    \end{tabular}
\end{table}

\subsubsection{Non-hyper-threaded regime}

    Table \ref{table:AWS-NHT} shows the results for AWS with Hyper-Threading disabled. Thus only 18 out of 36 cores were used to run the benchmark, and each core was able to boost into the turbo-frequency \cite{turbo-boost-technology}. It can be seen that the ratio of absolute speedup to the number of nodes still rapidly degrades with increased node count.

\begin{table}
    \caption{
        [AWS-NHT] Results for Amazon Web Services c4.8xlarge instances with hyper-threading disabled. Core count is given for physical (non-hyper-threaded) computing cores. Numbers of computing nodes (Nodes) and total computing cores (Cores) are given together with the maximum achieved (Rmax) and peak (Rpeak) performance indicators, and the absolute achieved speedup (Speedup).
    }
    \label{table:AWS-NHT}
    \begin{tabular}{ccccc}
        Nodes & Cores & Rmax (TFLOPS) & Rpeak (TFLOPS) & Speedup \\
        \hline
        1     & 18    & 0.64          & 0.82           & 1.00    \\
        2     & 36    & 1.14          & 1.63           & 1.77    \\
        4     & 72    & 1.94          & 3.26           & 3.02    \\
        8     & 144   & 3.51          & 6.53           & 5.47    \\
        16    & 288   & 5.59          & 13.05          & 8.71    \\
        32    & 576   & 10.68         & 26.10          & 16.65  
    \end{tabular}
\end{table}

\subsubsection{Non-hyper-threaded regime with placement groups}
    
    Table \ref{table:AWS-NHT-PG} shows the HPL benchmark results with hyper-threading disabled and placement group option enabled. A placement group is a logical grouping of instances within a single availability zone, recommended for applications that benefit from low network latency, high network throughput, or both \cite{placement-groups}. It can be seen, however, that the ratio of absolute speedup to the number of nodes shows marginal differences with respect to the previous scenario, where placement group option was not used.

\begin{table}
    \caption{
        [AWS-NHT-PG] Results for Amazon Web Services c4.8xlarge instances with hyper-threading disabled and with placement group option enabled at provision time. Core count is given for physical (non-hyper-threaded) computing cores. Numbers of computing nodes (Nodes) and total computing cores (Cores) are given together with the maximum achieved (Rmax) and peak (Rpeak) performance indicators, and the absolute achieved speedup (Speedup).
    }
    \label{table:AWS-NHT-PG}
    \begin{tabular}{ccccc}
        Nodes & Cores & Rmax (TFLOPS) & Rpeak (TFLOPS) & Speedup \\
        \hline
        1 & 18 & 0.62 & 0.82 & 1.00 \\
        2 & 36 & 1.14 & 1.63 & 1.82 \\
        4 & 72 & 1.97 & 3.26 & 3.15 \\
        8 & 144 & 3.51 & 6.53 & 5.61 \\
        16 & 288 & 5.70 & 13.05 & 9.12 \\
        32 & 576 & 10.74 & 26.10 & 17.18
    \end{tabular}
\end{table}

\subsection{Microsoft Azure}
\label{subsec:AZ}

\subsubsection{F-series}
    
    Table \ref{table:AZ-F} shows the HPL benchmark results running on Azure Standard\_F16 instances. Although the overall performance degradation with increased node count is evident, it appears to be less severe than for AWS. The bare performance is worse however.

\begin{table}
    \caption{
        [AZ-F] Results for Azure F-series instances. Core count is given for physical (non-hyper-threaded) computing cores. Numbers of computing nodes (Nodes) and total computing cores (Cores) are given together with the maximum achieved (Rmax) and peak (Rpeak) performance indicators, and the absolute achieved speedup (Speedup).
    }
    \label{table:AZ-F}
    \begin{tabular}{ccccc}
        Nodes & Cores & Rmax (TFLOPS) & Rpeak (TFLOPS) & Speedup \\
        \hline
        1 & 16 & 0.48 & 0.6 & 1.00 \\
        2 & 32 & 0.87 & 1.2 & 1.82 \\
        4 & 64 & 1.49 & 2.4 & 3.14 \\
        8 & 128 & 3.04 & 4.8 & 6.38 \\
        16 & 256 & 5.33 & 9.6 & 11.18 \\
        32 & 512 & 10.53 & 19.2 & 22.11
    \end{tabular}
\end{table}

\subsubsection{A-series}
Table \ref{table:AZ-F} shows the HPL benchmark results running on Azure Standard\_A9 instances using Infiniband interconnect network. The low-latency network interconnect definitely affects the scaling, increasing the speed-up ratio ~from 0.5 to ~0.9 for 32 compute nodes. The bare performance figures, however are still better for AWS due to the higher processor speed.

\begin{table}
    \caption{
    [AZ-A] Results for Azure A-series instances with Infiniband \cite{2001-pfister-infiniband} interconnect network. Core count is given for physical (non-hyper-threaded) computing cores. Numbers of computing nodes (Nodes) and total computing cores (Cores) are given together with the maximum achieved (Rmax) and peak (Rpeak) performance indicators, and the absolute achieved speedup (Speedup).
    }
    \label{table:AZ-A}
    \begin{tabular}{ccccc}
        Nodes & Cores & Rmax (TFLOPS) & Rpeak (TFLOPS) & Speedup \\
        \hline
        1 & 16 & 0.30 & 0.65 & 1.00 \\
        2 & 32 & 0.58 & 1.3 & 1.95 \\
        4 & 64 & 1.16 & 2.6 & 3.91 \\
        8 & 128 & 2.25 & 5.2 & 7.56 \\
        16 & 256 & 4.42 & 10.4 & 14.88 \\
        32 & 512 & 8.59 & 20.8 & 28.94
    \end{tabular}
\end{table}

\subsubsection{H-series}

    Table \ref{table:AZ-A} shows the HPL benchmark results running on Azure Standard\_H16r instances using Infiniband interconnect network. The low-latency network interconnect enables the best scaling pattern, with sustained ratio above 0.9 in the 1-32 node count (1-512 computing cores) range. The bare performance figures are best of all cases studied, even when compared with the traditional supercomputing system of reference.

\begin{table}
    \caption{
    [AZ-H] Results for Azure H-series instances with Infiniband \cite{2001-pfister-infiniband} interconnect network. Core count is given for physical (non-hyper-threaded) computing cores. Numbers of computing nodes (Nodes) and total computing cores (Cores) are given together with the maximum achieved (Rmax) and peak (Rpeak) performance indicators, and the absolute achieved speedup (Speedup).
    }
    \label{table:AZ-H}
    \begin{tabular}{ccccc}
        Nodes & Cores & Rmax (TFLOPS) & Rpeak (TFLOPS) & Speedup \\
        \hline
        1 & 16 & 0.61 & 0.8 & 1.00 \\
        2 & 32 & 1.22 & 1.6 & 2.01 \\
        4 & 64 & 2.40 & 3.2 & 3.93 \\
        8 & 128 & 4.69 & 6.4 & 7.69 \\
        16 & 256 & 9.09 & 12.8 & 14.91 \\
        32 & 512 & 17.26 & 25.6 & 28.33
    \end{tabular}
\end{table}

\subsection{Rackspace}
\label{subsec:RS}

    Table \ref{table:RS} shows the HPL benchmark results running on Rackspace Compute1-60 instances. Overall, the results are similar to AWS and Azure. A slight variation (spike) in the speedup ratio for 16 nodes can be associated with the underlying network topology of the cloud datacenter.

\begin{table}
    \caption{
    [RS] Results for Rackspace Compute1-60 instances. Core count is given for physical (non-hyper-threaded) computing cores. Numbers of computing nodes (Nodes) and total computing cores (Cores) are given together with the maximum achieved (Rmax) and peak (Rpeak) performance indicators, and the absolute achieved speedup (Speedup).
    }
    \label{table:RS}
    \begin{tabular}{ccccc}
        Nodes & Cores & Rmax (TFLOPS) & Rpeak (TFLOPS) & Speedup \\
        \hline
        1     & 32    & 0.16          & 0.7           & 1.00    \\
        2     & 64    & 0.28          & 1.4           & 1.68    \\
        4     & 128   & 0.57          & 2.8           & 3.46    \\
        8     & 256   & 0.98          & 5.6           & 5.97    \\
        16    & 512   & 2.14          & 11.2          & 13.07   \\
        32    & 1024  & 3.04          & 22.4          & 18.55  
    \end{tabular}
\end{table}

\subsection{IBM SoftLayer}
\label{subsec:SL}

    Table \ref{table:SL} shows the HPL benchmark results running on SoftLayer virtual servers. The network quickly saturates at scale, demonstrating the worst performance out of all cases studied. The processor clock speeds are also inferior when compared to other cloud options.

\begin{table}
    \caption{
    [SL] Results for SoftLayer virtual servers with 32 cores, 64 GB RAM and 1Gb/s bandwidth. Core count is given for physical (non-hyper-threaded) computing cores. Numbers of computing nodes (Nodes) and total computing cores (Cores) are given together with the maximum achieved (Rmax) and peak (Rpeak) performance indicators, and the absolute achieved speedup (Speedup).
    }
    \label{table:SL}
    \begin{tabular}{ccccc}
        Nodes & Cores & Rmax (TFLOPS) & Rpeak (TFLOPS) & Speedup \\
        \hline
        1     & 32    & 0.57          & 0.525         & 1.00    \\
        2     & 64    & 0.66          & 1.05          & 1.16    \\
        4     & 128   & 0.44          & 2.1           & 0.77    \\
        8     & 256   & 0.67          & 4.2           & 1.17    \\
        16    & 512   & 1.46          & 8.4           & 2.58    \\
        32    & 1024  & 2.46          & 16.8          & 4.33  
    \end{tabular}
\end{table}

\subsection{NERSC}
\label{subsec:NS}

    
    Table \ref{table:NERSC-E} shows the HPL benchmark results running on NERSC Edison supercomputer with hyper-threading enabled. Edison is a Cray XC30, with a peak performance of 2.57 PFLOPS, 133,824 compute cores, 357 terabytes of memory, and 7.56 petabytes of disk, holding number 60 rank on the top500 list of the best supercomputers at the moment of this writing \cite{top500-edison-supercomputer}.

\begin{table}
    \caption{
    [NERSC-E] Results for NERSC Edison supercomputer with hyper-threading enabled. Core count is given for virtual (hyper-threaded) computing cores. Numbers of computing nodes (Nodes) and total computing cores (Cores) are given together with the maximum achieved (Rmax) and peak (Rpeak) performance indicators, and the absolute achieved speedup (Speedup).
    }
    \label{table:NERSC-E}
    \begin{tabular}{ccccc}
        Nodes & Cores & Rmax (TFLOPS) & Rpeak (TFLOPS) & Speedup \\
        \hline
        1 & 48 & 0.38 & 0.9 & 1.00 \\
        2 & 96 & 0.73 & 1.8 & 1.91 \\
        4 & 192 & 1.34 & 3.6 & 3.48 \\
        8 & 384 & 2.79 & 7.2 & 7.27 \\
        16 & 768 & 5.40 & 14.4 & 14.06 \\
        32 & 1536 & 10.44 & 28.8 & 27.17
    \end{tabular}
\end{table}

\section{Discussion}
\label{sec:discussion}

\begin{figure}[ht!]
  \includegraphics[width = 0.48\textwidth]{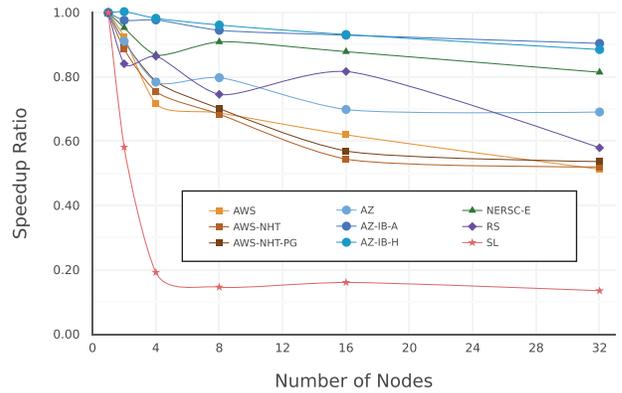}
  \caption{
    \label{fig-1}
    Speedup ratios (the ratios of maximum speedup Rmax to peak speedup Rpeak) against the number of nodes for all benchmarked cases. Speedup ratio for 1,2,4,8,16 and 32 nodes are investigated and given by points. Lines are drawn to guide the eye. The legend is as follows: AWS - Amazon Web Services in the default hyper-threaded regime; AWS-NHT - same, with hyperthreading disabled; AWS-NHT-PG - same, with placement group option enabled; AZ - Microsoft Azure standard F16 instances; AZ-IB-A - same provider, A9 instances; AZ-IB-H - same provider, H16 instances; RS - Rackspace compute1-60 instances; SL - IBM/Softlayer virtual servers; NERSC -  Edison computing facility of the National Energy Research Scientific Computing Center.
    }
\end{figure}

    In Fig.~\ref{fig-1} we present a comparison of the speedup ratios for the scenarios described the previous part. As it can be seen, Microsoft Azure outperforms other cloud providers because of the low latency interconnect network that facilitates efficient scaling. SoftLayer has the least favorable speedup ratio at scale, likely because of the interconnect network again. AWS and Rackspace show a significant degree of parallel performance degradation, such that at 32 nodes the measured performance is about one-half of the peak value.

\begin{figure}[ht!]
  \includegraphics[width = 0.48\textwidth]{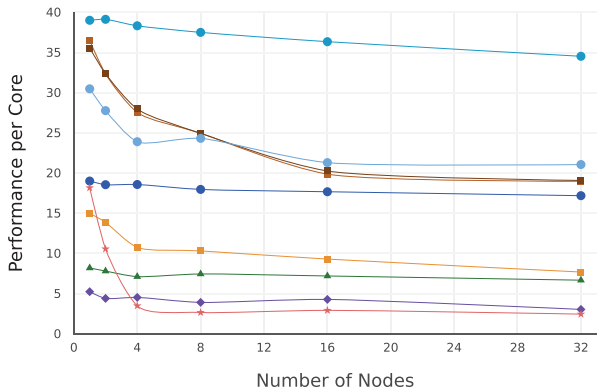}
  \caption{
    \label{fig-2}
    Performance per core in giga-FLOPS against the number of nodes for all benchmarked cases. Performance per core is obtained by dividing the maximum performance by the total number of computing cores. The legend is the same as in Fig. \ref{fig-1}. Lines are given to guide the eye.
  }
\end{figure}

    Fig.~\ref{fig-2} shows a comparative plot of the performance per core in giga-FLOPS for the previously described scenarios. Microsoft Azure H-instances are the highest performing option in this view as well (AZ-IB-H). One interesting fact is that although Microsoft Azure A-instances (AZ-IB-A) show better overall scaling in Fig.~\ref{fig-1}, AWS c4.8xlarge instances deliver better performance per core for up to 16 nodes. This is likely because of faster processors speed. NERSC Edison supercomputer delivers a rather low performance per core metric, likely due to the type of processors used.
    
    Our results demonstrate that the current generation of publicly available cloud computing systems are capable of delivering comparable, if not better, performance than the top-tier traditional high performance computing systems. This fact confirms that cloud computing is already a viable and cost-effective alternative to traditional cost-intensive supercomputing procurement. We believe that with further advancements in virtualization, such as low-overhead container technology, and future improvements in cloud datacenter hardware we may experience a large-scale migration from on-premises to cloud-based usage for high performance applications, similar to what happened with less compute-intensive workloads.

\section{Conclusion}
\label{sec:conclusion}

    We benchmarked the performance of the best available computing hardware from public cloud providers with high performance Linpack. We optimized the benchmark for each computing environment and evaluated the relative performance for distributed memory calculations. We found Microsoft Azure to deliver the best results, and demonstrated that the performance per single computing core on public cloud to be comparable to modern traditional supercomputing systems. Based on our findings we suggest that the concept of high performance computing in the cloud is ready for a widespread adoption and can provide a viable and cost-efficient alternative to capital-intensive on-premises hardware deployments.

\section{Acknowledgement}
\label{sec:acknowledgement}

    Authors would like to thank Michael G. Haverty for reading the manuscript, and acknowledge support from the National Energy Research Scientific Computing Center in a form of a startup allocation.

\bibliographystyle{IEEEtran}
\bibliography{ref}

\end{document}